\documentclass[aps, nofootinbib,reprint]{revtex4-1}
\usepackage{amsmath}
\usepackage{graphicx}
\usepackage{epsfig}
\usepackage{amssymb}
\usepackage{enumitem}

\newcommand{\ii}{\mathrm{i}}

\def\diff{d}
\def\dd{d}
\def\nn{\nonumber}

\begin{document}

\title{The gravity dual of supersymmetric gauge theories on a squashed five-sphere}

\author{Luis F. Alday, Martin Fluder, Paul Richmond and James Sparks}
\email[Email:]{alday, fluder, richmond, sparks@maths.ox.ac.uk}
\vskip.2in
\affiliation{Mathematical Institute, University of Oxford, Woodstock Road, Oxford, OX2 6GG, UK}
 
\begin{abstract}
\noindent We present the gravity dual of large $N$ supersymmetric gauge theories on a squashed five-sphere. The one-parameter family of solutions is constructed in Euclidean Romans $F(4)$ gauged supergravity in six dimensions, and uplifts to massive type $IIA$ supergravity. By renormalizing the theory with appropriate counterterms we evaluate the renormalized on-shell action for the solutions. We also evaluate the large $N$ limit of the gauge theory partition function, and find precise agreement.  
\end{abstract}

\maketitle

\section{Supersymmetric gauge theories on a squashed five-sphere}\label{sec1}

In \cite{Imamura:2012xg} supersymmetric gauge theories with general matter content were defined on the $SU(3) \times U(1)$ symmetric squashed five-sphere. The background metric is
\begin{eqnarray}
\label{5dmetric}
ds^2_5&=&\frac{1}{s^2}(d \tau+C)^2+ d \sigma^2 + \frac{1}{4}\sin^2\sigma(d \theta^2+\sin^2\theta d \varphi^2) \nonumber\\
& &+\frac{1}{4}\cos^2\sigma\sin^2\sigma (d \psi+\cos\theta d\varphi)^2~,
\end{eqnarray}
where $C=-\frac{1}{2}\sin^2 \sigma(d\psi+\cos \theta d\varphi)$ and $s$ is the squashing parameter. The round 
sphere corresponds to  $s=1$. 
The theory preserves $3/4$ of the supersymmetry of the round sphere, provided one turns on a background $SU(2)_R$ gauge field
\begin{eqnarray}
\label{AR}
A^R &=& \frac{(1+ \sqrt{1-s^2})\sqrt{1-s^2}}{s^2} (d \tau+C)~,
\end{eqnarray}
where we have embedded $U(1)_R \subset SU(2)_R$. The background then admits a Killing spinor that solves 
the  Killing spinor 
equation in \cite{Imamura:2012xg} and transforms in the ${\bf 3}$ of $SU(3)$. The perturbative partition function of the gauge theories was 
 computed in \cite{Imamura:2012bm} (see also \cite{Lockhart:2012vp}) and the final formula involves triple sine functions, generalizing the double 
 sine functions that appear for squashed three-spheres \cite{Hama:2011ea}. 

A particular class of five-dimensional gauge theories, with gauge group $USp(2N)$ and arising from a $D4-D8$-system, is expected to have a large $N$ description in terms of massive type $IIA$ supergravity \cite{Ferrara:1998gv,Brandhuber:1999np}.  In \cite{Jafferis:2012iv} the large $N$ limit of the partition function of these theories on the {\it round} sphere was computed and successfully compared to the entanglement entropy of the dual warped $AdS_6 \times S^4$ supergravity solution. 

One can compute the large $N$ limit of the $USp(2N)$ gauge theory partition function $Z_s$ for the squashed background (\ref{5dmetric}), (\ref{AR}). The corresponding free energy ${\cal F}_s = - \log Z_s$ is given by
\begin{equation}
\label{free}
{\cal F}_s =  \frac{1}{27s^2} \frac{\left(3-\sqrt{1-s^2}\right)^3}{1-\sqrt{1-s^2}} ~{\cal F}_1~.
\end{equation}
Here ${\cal F}_1$ is the free energy on the round sphere, which scales as $N^{5/2}$ \cite{Brandhuber:1999np,Jafferis:2012iv}. 
The computation of (\ref{free}) involves asymptotic expansions of the triple sine function and standard large $N$ matrix model techniques, and details will appear in \cite{Alday:2014bta}. Similarly, we have computed the large $N$ limit of the VEV of a BPS Wilson loop wrapping the 
$\tau$ circle at $\sigma=0$, finding
\begin{equation}
\label{wilson}
\log\, \langle W\rangle_s =  \frac{3-\sqrt{1-s^2}}{3(1+\sqrt{1-s^2})}~\log\, \langle W\rangle_1~,
\end{equation}
where $\log \, \langle W\rangle_1$ scales as $N^{1/2}$ \cite{Assel:2012nf}.

In the remainder of this letter we will reproduce (\ref{free}) and (\ref{wilson}) from a dual supergravity computation. 

\section{Euclidean Romans supergravity}\label{sec2}

In order to find supergravity duals of the above theories put on general background five-manifolds it is natural to work in the six-dimensional Romans $F(4)$ supergravity theory \cite{Romans:1985tw}. The key here is that, as shown in \cite{Cvetic:1999un}, the Romans theory is a consistent truncation of massive type $IIA$ supergravity on $S^4$. In particular, the $AdS_6$ vacuum uplifts to the warped $AdS_6 \times S^4$ solution mentioned above, relevant for the round five-sphere. The bosonic fields  consist of the metric, a dilaton $\phi$, a two-form 
potential $B$, a one-form potential $A$, together with an $SO(3) \sim SU(2)$ gauge field $A^i$, $i=1,2,3$. 
It is convenient to introduce the scalar field $X\equiv \exp(-\phi/2\sqrt{2})$, and define the field strengths $H=d B$, $F =d A + \frac{2}{3} B$, $F^i = d A^i-\frac{1}{2}\epsilon_{ijk}A^j\wedge A^k$, where without loss of generality we have set the gauge coupling to 1. 

The equations of motion for the Romans theory in Lorentz signature appear in 
 \cite{Romans:1985tw, Cvetic:1999un}. However, in order to compute
 the holographic free energy we will work in Euclidean signature. This Wick rotation
is not entirely 
straightforward due to Chern-Simons-type couplings. The Euclidean equations of motion are
\cite{Alday:2014bta}
\begin{eqnarray}
\label{FullEOM}
\diff\left(X^4 * H\right) &=& \tfrac{\ii}{2}F\wedge F + \tfrac{\ii}{2}F^i\wedge F^i +\tfrac{2}{3}  X^{-2}*F~,\nn\\
\diff(X^{-2}*F) &=& - \ii F\wedge H~, \nn \\
D(X^{-2}*F^i) & = & - \ii F^i\wedge H~,\nn\\
\diff\left(X^{-1}*\dd X\right) &=& -  \left(\tfrac{1}{6}X^{-6}-\tfrac{2}{3}X^{-2}+\tfrac{1}{2}X^2\right)*1 \nn \\
&&-\tfrac{1}{8}X^{-2}\left(F\wedge *F+F^i\wedge *F^i\right) \nn \\
& &+ \tfrac{1}{4}X^4H\wedge *H ~.
\end{eqnarray}
Here $D\omega^i=\dd\omega^i -\epsilon_{ijk}A^j\wedge \omega^k$ is the 
$SO(3)$ covariant derivative. 
Finally, the Einstein equation is
\begin{eqnarray}
R_{\mu\nu} &=& 4X^{-2}\partial_\mu X\partial_\nu X + \left(\tfrac{1}{18}X^{-6}-\tfrac{2}{3}X^{-2}-\tfrac{1}{2}X^2\right) g_{\mu\nu} \nn \\
& &+ \tfrac{1}{4}X^4\left(H^2_{\mu\nu}-\tfrac{1}{6}H^2g_{\mu\nu}\right) \nn
 +  \tfrac{1}{2}X^{-2}\left(F^2_{\mu\nu}-\tfrac{1}{8}F^2g_{\mu\nu}\right) \\
 & &+  \tfrac{1}{2}X^{-2}\left((F^i)^2_{\mu\nu}-\tfrac{1}{8}(F^i)^2g_{\mu\nu}\right)~,
\end{eqnarray}
where $F^2_{\mu\nu} = F_{\mu\rho} F_\nu{}^\rho$, $H^2_{\mu\nu}=H_{\mu\rho\sigma}H_{\nu}^{\ \rho\sigma}$.

A solution to the above equations of motion is 
supersymmetric provided the Killing spinor equation and dilatino 
equation hold \cite{Alday:2014bta}:
\begin{eqnarray}
D_\mu \epsilon_I & =&  \frac{\ii}{4\sqrt{2}} ( X + \tfrac{1}{3} X^{-3} ) \gamma_\mu \gamma_7 \epsilon_I - \frac{1}{48} X^2 H^{\nu\rho\sigma} \gamma_{\nu\rho\sigma} \gamma_\mu \gamma_7 \epsilon_I   \nn \\
&&- \frac{\ii}{16\sqrt{2}} X^{-1} F_{\nu\rho} ( \gamma_\mu{}^{\nu\rho} - 6 \delta_\mu{}^\nu \gamma^\rho ) \epsilon_I \nn\\
& & + \frac{1}{16\sqrt{2}}X^{-1} F_{\nu\rho}^i ( \gamma_\mu{}^{\nu\rho} - 6 \delta_\mu{}^\nu \gamma^\rho ) \gamma_7 ( \sigma^i )_I{}^J \epsilon_J ~,  \\
 0 &=& - \ii X^{-1} \partial_\mu X \gamma^\mu \epsilon_ I + \frac{1}{2\sqrt{2}}  \left( X - X^{-3} \right) \gamma_7 \epsilon_I \nn \\
 & &+ \frac{\ii}{24} X^2 H_{\mu\nu\rho} \gamma^{\mu\nu\rho} \gamma_7 \epsilon_I \nonumber - \frac{1}{8\sqrt{2}} X^{-1} F_{\mu\nu} \gamma^{\mu\nu} \epsilon_I \\
 & &- \frac{\ii}{8\sqrt{2}} X^{-1} F^i_{\mu\nu} \gamma^{\mu\nu} \gamma_7 ( \sigma^i )_I{}^J \epsilon_J~.
\end{eqnarray}
Here $\epsilon_I$, $I=1,2$, are two Dirac spinors, $\gamma_\mu$ generate the Clifford 
algebra $\mathrm{Cliff}(6,0)$ in an orthonormal frame, and we have defined the chirality operator
 $\gamma_7 = \ii \gamma_{123456}$, which satisfies $\gamma_7^2=1$. 
The $SO(3) \sim SU(2)$ gauge field $A^i$ is an R-symmetry gauge field, 
 with the spinor $\epsilon_I$ transforming in the two-dimensional representation via
  the Pauli matrices $(\sigma^i)_I{}^J$. Thus the covariant derivative acting on the spinor is $D_\mu\epsilon_I=\nabla_\mu\epsilon_I+\frac{\ii}{2} A_\mu^i(\sigma^i)_I{}^J\epsilon_J$.

The theory possesses a gauge invariance 
$A\rightarrow A+\frac{2}{3}\lambda$, $B\rightarrow B-\diff\lambda$, where 
$\lambda$ is any one-form. Using this freedom we
fix the gauge $A=0$, leaving $F=\frac{2}{3}B$: the $B$-field ``eats'' the $U(1)$ gauge field $A$ in a Higgs-like mechanism.

\section{The solution}

The squashed five-sphere background of section \ref{sec1} has $SU(3) \times U(1)$ symmetry. One expects this symmetry to be preserved by the bulk solution. This leads to the following ansatz for the supergravity fields 
\begin{eqnarray}
ds^2_6 &=& \alpha^2(r)dr^2+\gamma^2(r)(d\tau+C)^2+\beta^2(r)\Big[d \sigma^2 \nn\\
&&+ \frac{1}{4}\sin^2\sigma(d \theta^2+\sin^2\theta d \varphi^2)\nn\\
&&+\frac{1}{4}\cos^2\sigma\sin^2\sigma (d \psi+\cos\theta d\varphi)^2\Big]~,\nn\\
B&=& p(r)dr\wedge (d\tau+C) + \frac{1}{2}q(r)dC~,\nn\\
A^i &=& f^i(r)(d\tau +C)~,
\end{eqnarray}
where also $X=X(r)$. We have constructed a smooth, supersymmetric, asymptotically locally Euclidean AdS solution 
to the equations in section \ref{sec2}, which has as conformal boundary the squashed five-sphere 
background of section \ref{sec1}. The function $\beta(r)$ can be set to its $AdS_6$ value by
using reparametrization invariance, $\beta(r)= {3\sqrt{6r^2-1}}/{\sqrt{2}}$.
%
%\begin{equation}
%\beta(r)= \frac{3\sqrt{6r^2-1}}{\sqrt{2}} ~.
%\end{equation}
%
Furthermore, we have performed an $SO(3)\sim SU(2)$ rotation so as to set $f^1(r)=f^2(r)=0$, and renamed $f^3(r) \equiv f(r)$. Even though we are not able to give a closed expression for the solution, it is possible to give it as an expansion around different limits. 

\subsection{Expansion around the conformal boundary}
Finding the gravity dual to a theory on a prescribed conformal boundary may be regarded as a filling problem 
in supergravity. As such, it is natural to solve the supergravity equations order by order in an expansion around the 
boundary at $r=\infty$.
We have computed this expansion  up to order ${\cal O}(1/r^9)$. The first terms are given by
\begin{widetext}
\begin{eqnarray}
\alpha(r)&=&\frac{3}{\sqrt{2}} \frac{1}{r}+\frac{8+s^2}{36 \sqrt{2} s^2}\frac{1}{r^3}+\ldots,~~~
\gamma(r)=\frac{3 \sqrt{3}}{s} r+\frac{-16+7 s^2}{12 \sqrt{3} s^3 }\frac{1}{r}-\frac{-1280+1120 s^2+241 s^4}{2592 \sqrt{3} s^5}\frac{1}{r^3}+\ldots,\nn \\
X(r)&=&1+\frac{1-s^2-3  \sqrt{1-s^2}}{54 s^2}\frac{1}{r^2}+\frac{s^2 \sqrt{1-s^2} \kappa}{12 \left(1- s^2+\sqrt{1-s^2}\right)}\frac{1}{r^3}+\ldots,\nn \\
 p(r) &=& -\frac{ \ii \sqrt{\frac{2}{3}} \left(s^2+3\sqrt{1-s^2}-1\right)}{s^3}\frac{1}{r^2}+\ldots,~~q(r) = -\frac{3 \ii \left(\sqrt{6} \sqrt{1-s^2}\right)}{s} r 
 + \frac{\sqrt{\frac{2}{3}} \ii \sqrt{1-s^2} \left(5 s^2+9  \sqrt{1-s^2}-5\right)}{3 s^3}\frac{1}{r}+\ldots,\nn\\
f(r) &=& \frac{1-s^2+\sqrt{1-s^2}}{s^2} +\frac{2 \left(-2+2 s^2- (2+s^2) \sqrt{1-s^2}\right)}{9 s^4}\frac{1}{r^2}+\frac{\kappa}{r^3}+\ldots~.
\end{eqnarray}
\end{widetext}
Notice that the squashing parameter $s$ arises as the boundary value $\lim_{r\rightarrow\infty}\gamma(r)/3\sqrt{3}r=s^{-1}$. In the limit $s=1$ the solution collapses to Euclidean $AdS_6$. The whole solution depends on the single parameter $s$. The extra parameter $\kappa$ is fixed by requiring a non-singular solution at the origin $r=1/\sqrt{6}$. Alternatively, this can be computed as an expansion, as is done at the end of next subsection. 

\subsection{Expansion around Euclidean $AdS_6$}

The solution presented in this letter is continuously connected to Euclidean $AdS_6$. Hence it can be given as a perturbation around this background. It is convenient to use the real expansion parameter $\delta$, related to the squashing parameter by
\begin{equation}
\tfrac{1}{s}=1+\delta^2~.
\end{equation}
We have explicitly computed the solution up to sixth order in $\delta$. At leading order we find
\begin{widetext}
\begin{eqnarray}
\alpha(r)&=&\frac{3 \sqrt{3}}{\sqrt{6 r^2-1}}+\frac{\left(-5 \sqrt{6}+330 \sqrt{6} r^2-3744 r^3+1620 \sqrt{6} r^4+8640 r^5-7560 \sqrt{6} r^6+5184 \sqrt{6} r^8\right)}{9 \sqrt{2} r^2 \left(6 r^2-1\right)^{9/2}} \delta ^2+\ldots,\nn\\
\gamma(r)&=&\frac{3 \sqrt{6 r^2-1}}{\sqrt{2}}-\frac{\left(55 \sqrt{2}-384 \sqrt{3} r+1080 \sqrt{2} r^2+768 \sqrt{3} r^3-5400 \sqrt{2} r^4+11232 \sqrt{2} r^6-11664 \sqrt{2} r^8\right) }{6 \left(6 r^2-1\right)^{7/2}}\delta ^2+\ldots,\nn\\
X(r) &=& 1-\frac{\left(\sqrt{2} \left(1-2 \sqrt{6} r+6 r^2\right)\right)}{3 \left(6 r^2-1\right)^2}\delta +\ldots,~~~~~~
q(r) = -\frac{3 i \sqrt{2} \left(-4+9 \sqrt{6} r-24 r^2-12 \sqrt{6} r^3+36 \sqrt{6} r^5\right) \delta }{\left(6 r^2-1\right)^2}+\ldots,\nn \\
p(r) &=& \frac{18 i \sqrt{2} \left(\sqrt{6}-16 r+12 \sqrt{6} r^2-12 \sqrt{6} r^4\right) }{\left(6 r^2-1\right)^3} \delta+\ldots,~~~
f(r) = \frac{\sqrt{2} \left(-3+8 \sqrt{6} r-36 r^2+36 r^4\right)}{\left(6 r^2-1\right)^2} \delta +\ldots,
\end{eqnarray}
\end{widetext}
One can explicitly check that each term of the solution above is non-singular at the origin $r=1/\sqrt{6}$, giving a regular solution on a manifold $M_6$ with the topology of a six-ball. 
 By comparing the two expansions we find
\begin{equation}
\frac{3\sqrt{3}}{4} \kappa=\delta +\frac{\sqrt{2}}{3}\delta ^2+\frac{113 }{36}\delta ^3+\frac{25}{9 \sqrt{2}} \delta ^4+\frac{1127}{288}\delta ^5+\frac{35 }{9 \sqrt{2}}\delta ^6+...
\end{equation}
which is used in evaluating the on-shell action below.

\section{Comparison}
The bulk supergravity action of the Romans theory, in Euclidean signature in the gauge $A=0$, is
\begin{eqnarray}
S_{\mathrm{bulk}} & =  & -\frac{1}{16\pi G_N}\int_{M_6} \Big[R*1-4X^{-2}\diff X\wedge *\diff X \nn\\
&& -\left(\tfrac{2}{9}X^{-6}-\tfrac{8}{3}X^{-2}-2X^2\right)*1\nn\\
&&-\tfrac{1}{2}X^{-2} \left(\tfrac{4}{9}B\wedge *B+ F^i \wedge * F^i\right) -\tfrac{1}{2}X^4H\wedge *H \nn\\
&&- \ii B \wedge \big( \tfrac{2}{27}  B\wedge B + \tfrac{1}{2} F^i \wedge F^i \big)\Big]~. \label{Fullaction}
\end{eqnarray}
Here $G_N$ is the six-dimensional Newton constant. More precisely, we should cut off the manifold $M_6$ at 
some large constant radius $r=\rho$, and include the Gibbons-Hawking boundary term
\begin{eqnarray}
S_{\mathrm{GH}} &=& -\frac{1}{8\pi G_N}\int_{\partial M_6} K\sqrt{\det h}\, \dd^5 x~,
\end{eqnarray}
where $h_{ij}$ is the induced metric on the boundary $\partial M_6=\{r=\rho\}\cong S^5$, and 
$K$ denotes the trace of the second fundamental form. The total action 
is divergent as one sends $\rho\rightarrow\infty$, but may be regularized 
using holographic renormalization techniques. This leads to the following boundary 
counterterms \cite{Alday:2014bta}
\begin{eqnarray}
S_{\mathrm{ct}}&=&  \frac{1}{8\pi G_N}\int_{\partial M_6}\Bigg\{\Big[\frac{4\sqrt{2}}{3}+\frac{1}{2\sqrt{2}}R(h)-\frac{1}{6\sqrt{2}}\|B\|^2_h\nn\\
&&+\frac{3}{4\sqrt{2}}R(h)_{ij}R(h)^{ij}-\frac{15}{64\sqrt{2}}R(h)^2-\frac{3}{4\sqrt{2}}\|F^i\|^2_h\nn\\
&&+\frac{1}{12\sqrt{2}}\mathrm{Tr}_{h}B^4+\frac{5}{8\sqrt{2}}\|\diff *_{h}B+\frac{\ii\sqrt{2}}{3}B\wedge B\|^2_{h}\nn\\
&&-\frac{1}{4\sqrt{2}}\langle B,\diff\delta_{h}B+\frac{\ii \sqrt{2}}{3}\dd*_{h}B\wedge B\rangle_{h}-\frac{1}{\sqrt{2}}\|\dd B\|^2_{h}\nn\\
&&+\frac{4\sqrt{2}}{3}(1-X)^2-
\frac{1}{\sqrt{2}}\langle \mathrm{Ric}(h)\circ B,B\rangle_{h}\nn\\
&&+\frac{9}{32\sqrt{2}}R(h)\|B\|^2_{h}
-\frac{13}{192\sqrt{2}}\|B\|^4_{h} 
\Big]\sqrt{\det h}\,\dd^5x\nn\\
&&-\frac{1}{4\sqrt{2}}B\wedge \big[\diff *_h\diff B + \frac{\sqrt{2}\ii}{3}B\wedge\delta_h B\nn\\
&& - \frac{2}{9}B\wedge *_{h}(B\wedge B)\big]\Bigg\}~.
\end{eqnarray}
Here $\mathrm{Ric}(h)_{ij}=R(h)_{ij}$ denotes the Ricci tensor of the metric $h_{ij}$, with $R(h)$ the 
Ricci scalar. The inner product of two $p$-forms $\nu_1$, $\nu_2$ is defined by 
$\langle \nu_1,\nu_2\rangle_h\sqrt{\det h}\, \diff^5x = \nu_1\wedge *_h\nu_2$, which then 
also defines the square norm via $\|\nu\|_h^2=\langle \nu,\nu\rangle_h$.
The adjoint $\delta_h$ of $\diff$ with respect to $h_{ij}$ acting on the two-form $B$ is $\delta_hB=*_h\diff *_hB$, 
and we have also defined $\mathrm{Tr}_{h}B^4 \equiv B_i^{\ j}B_j^{\ k}B_k^{\ l}B_l^{\ i}$. 
Finally, we have defined the $p$-form $ (S\circ\nu)_{i_1\cdots i_p} \equiv  {S_{[i_1}}^{j}\nu_{|j|i_2\cdots i_p]}$, 
where $S_{ij}$ is any symmetric 2-tensor, and $\nu$ is any $p$-form.

Adding the contributions and taking the cut-off to infinity we obtain
\begin{eqnarray}
S_{\mathrm{bulk}}+S_{\mathrm{GH}}+S_{\mathrm{ct}
}&=& -\frac{27 \pi^2}{4 G_N} \left( 1+\frac{8}{3}  \delta^2 +\frac{16 \sqrt{2}}{27} \delta^3+\frac{68 }{27}\delta^4 \right.\nn \\
& &\left. +\frac{28 \sqrt{2}}{27} \delta ^5+\frac{32 }{27}\delta ^6 +\ldots\right)~.
\end{eqnarray}
This should be identified with the holographic free energy. Recalling that $s^{-1}=1+\delta^2$, this precisely agrees with (\ref{free}) to sixth order in $\delta$. It should be straightforward to extend this agreement to higher orders. 

The BPS Wilson loop (\ref{wilson}) maps to a fundamental string in type $IIA$, at the ``pole'' of the internal $S^4$ \cite{Assel:2012nf}. 
The string wraps the  surface $\Sigma$ spanned by the $\tau$ and $r$ directions at $\sigma=0$. The renormalized string action is
\begin{equation}
S_{\mathrm{string}} = \int_{\Sigma}\left[X^{-2}\sqrt{\det \gamma}\, d^2x + \ii B\right]-\frac{3}{\sqrt{2}}\mathrm{length}(\partial\Sigma)~,
\end{equation}
where $\gamma_{ab}$ is the induced metric and
the second term is a boundary counterterm. We may evaluate this up to sixth order in $\delta$ for our solution to obtain
\begin{eqnarray}
S_{\mathrm{string}} &= &\left(1-\frac{4 \sqrt{2} \delta }{3}+\frac{8 \delta ^2}{3} -\frac{5 \sqrt{2} \delta ^3}{3}+\frac{4 \delta ^4}{3}\right.\nn\\
&&\left.-\frac{7 \delta ^5}{12 \sqrt{2}}+ 0~ \delta^6+\ldots\right)~ S_{\mathrm{string}}\mid_{\delta=0}~,
\end{eqnarray}
which precisely matches (\ref{wilson}).

\vspace{-0.1in}

\section{A conjecture}
A supersymmetric solution admits an $SU(2)$ doublet of Killing spinors $\epsilon_I$. Provided the Killing spinor satisfies a symplectic Majorana condition ${\cal C} \epsilon_I^* = \varepsilon_I^{~J} \epsilon_J$, where ${\cal C}$ is the charge conjugation matrix (${\cal C}^{-1} \gamma_\mu {\cal C}=\gamma_\mu^*$), it can be shown \cite{Alday:2014bta} that 
\begin{equation}
\label{killing}
K_\mu = \varepsilon^{IJ}\epsilon_I^T \mathcal{C} \gamma_\mu \epsilon_J
\end{equation}
is a real Killing vector.
For our solution the Killing spinor has 3 integration constants, corresponding to the fact that it is 3/4 BPS, and for an appropriate 
choice of the Killing spinor in (\ref{killing}) we obtain
\begin{equation}
\label{killingb}
K=b_1 \partial_{\varphi_1}+b_2 \partial_{\varphi_2} +b_3 \partial_{\varphi_3}~,
\end{equation}
where 
\begin{equation}
b_1=1+\sqrt{1-s^2},~~~b_2=b_3=1-\sqrt{1-s^2}~,
\end{equation}
and $\varphi_{1},\varphi_2,\varphi_3$ are the standard $2\pi$ periodic azimuthal variables
$\varphi_1=-\tau$, $\varphi_2=\tau-\frac{1}{2}(\psi+\varphi)$, $\varphi_3=\tau-\frac{1}{2}\left(\psi-\varphi \right)$, 
%
%\begin{equation}
%\varphi_1=-\tau,~~~\varphi_2=\tau-\frac{1}{2}(\psi+\varphi),~~~\varphi_3=\tau-\frac{1}{2}\left(\psi-\varphi \right)~,
%\end{equation}
%
embedding $S^5 \subset \mathbb{R}^2 \oplus  \mathbb{R}^2 \oplus  \mathbb{R}^2$.
Note that the large  $N$ free energy (\ref{free}) can then be written as
\begin{equation}
\label{freeb}
{\cal F}=\frac{\left( b_1+b_2+b_3 \right)^3}{27 b_1 b_2 b_3}{\cal F}_{\mathrm{round}}~. 
\end{equation}

\smallskip

It is then natural to conjecture 

\smallskip

\begin{enumerate}[leftmargin=0.5cm]
\item For any supersymmetric supergravity solution with the topology of the six-ball, with at least $U(1)^3$ isometry, and for which the Killing vector  (\ref{killing}) takes the form (\ref{killingb}), the holographic free energy is equal to (\ref{freeb}).
\item If we define a supersymmetric gauge theory on the conformal boundary of the background in point 1, the finite $N$ partition function depends only on $b_1,b_2,b_3$.
\end{enumerate}
These conjectures extend to $5d/6d$ the results proven for the analogous $3d/4d$ context in \cite{Alday:2013lba,Farquet:2014kma}. Conjecture 1 also extends to the BPS Wilson loop wrapping $\varphi_i$, at the origin of the perpendicular $\mathbb{R}^4$. In this case $\log\, \langle W\rangle_s =  \frac{(b_1+b_2+b_3)}{3 b_i}~\log\, \langle W\rangle_1$. 

In \cite{Alday:2014bta} we construct further families of supersymmetric backgrounds satisfying the conditions of point 1 and verify the conjecture for these cases. These include a supersymmetric solution with the $SU(2)$ gauge field turned off, with a squashing parameter but for which $b_i=1$. For this case the free energy does not depend on the squashing parameter, in full agreement with conjecture 1. 

\bigskip

\noindent 
 The work of L.~F.~A, M.~F. and P.~R. is supported by ERC STG grant 306260. L.~F.~A. is a Wolfson Royal Society Research Merit Award holder.  J.~F.~S. is supported by a Royal Society University Research Fellowship.

\end{document}